\documentclass[twocolumn]{aastex6}

\epsscale{1.17} 

\usepackage{epsfig}
\usepackage{amsmath}
\usepackage{amssymb}
\usepackage{epstopdf}

\begin{document}

\title{Special supernova signature from BH-NS/BH progenitor systems}

\author{He Gao$^{1,*}$, Liang-Duan Liu$^{1}$, Wei-Hua Lei$^{2}$ and Litao Zhao$^{1}$}
\affiliation{
$^1$Department of Astronomy, Beijing Normal University, Beijing 100875, China; gaohe@bnu.edu.cn\\
$^2$Department of Astronomy, School of Physics, Huazhong University of Science and Technology, Wuhan, China.}

\begin{abstract}

The gravitational-wave detection by the LIGO-Virgo scientific collaboration shows that the black hole and neutron star (BH-NS) or BH-BH systems with a BH mass of tens of solar masses widely exist in the universe.  Two main types of scenarios have been invoked for the formation of BH-NS/BH systems, including isolated binary evolution in galactic fields and dynamical interactions in dense environments. 
Here we propose that if the BH-NS/BH systems are formed from isolated binary evolution, the supernova (SN) signal associated with the second core collapse would show some identifiable features, due to the accretion feedback from the companion BH. Depending on the binary properties, we show that the SN lightcurve could present a sharp peak around $\sim10$ days, with luminosity even at the level of the super luminous SNe ( e.g. $\sim10^{44}~\rm erg~s^{-1}$) or present a plateau feature lasting for several tens of days with regular luminosity of core collapse SNe. Comparing the event rate density of these special SN signals with the event rate density of LIGO-Virgo detected BH-NS/BH systems could help to distinguish the BH-NS/BH formation channel.

\end{abstract}


\section{Introduction}
With the first detection of binary black hole (BBH) coalescence signal, GW150914~\citep{GW150914}, the LIGO-Virgo scientific collaboration (LVC) opened up the field of gravitational-wave astrophysics. In the first two observing runs (O1 and O2), LVC discovered another nine BBH merger events, providing an estimate of the BBH merger rate density $R=53.2_{-28.2}^{+55.8}~{\rm Gpc^{-3}~yr^{-1}}$\citep{GWT1}. During the third observing run (O3), more BBH merger events were continued to be discovered with some interesting special cases, such as coalescence with asymmetric masses (e.g., GW190412, see \cite{GW190412}; and GW190814, see \cite{GW190814}). Nevertheless, some black hole and neutron star (BH-NS) candidate events at various confidence levels have also been discovered \citep{Anand20}.

The formation channel of BH-NS/BH systems is still under debated. Two main types of scenarios have been invoked for the formation of BH-NS/BH systems, including isolated binary evolution in galactic fields \citep{tutukov73,lipunov97,belczynski16} and dynamical interactions in dense environments \citep{sigurdsson93,portegies00,rodriguez15}. In the binary evolution scenario, the faster evolving star first produced a BH through core collapse, forming a BH + massive star binary. After a certain time delay, the second core collapse event would lead to a BH-NS/BH system formation. Two core collapses are likely to be accompanied by two supernova (SN) explosions.

In this Letter, we propose that the second SN explosion will be special because it has a very close black hole as its companion star. When the SN material expands and approaches to the companion BH, a violent accretion process could trigger strong feedback to the SN explosion. Here we show that once the feedback is energetic enough, the second SN would present some identifiable signatures. Comparing the event rate density of these special SN signals with the event rate density of LIGO-Virgo detected BH-NS/BH systems could help to distinguish the BH-NS/BH formation channel.

\section{Model Description} 
\label{Sec:Model}

Consider a binary system with a massive star and a companion BH (with a mass $M_{\mathrm{BH}}$), where the orbital separation is $d$. When the massive star explodes as an SN, a total mass of $M_{\rm ej}$ could be ejected with an explosion energy $E_{\mathrm{sn}}$. Based on numerical simulations of SN explosions, the density profile of SN ejecta could be described by a broken power law \citep{matzner99}
\begin{equation}
  \rho_{\mathrm{ej}} (v, t) = \left\{ \begin{array}{ll}
    \zeta_{\rho} \frac{M_{\mathrm{ej}}}{v_{\mathrm{tr}}^3 t^3} \left(
    \frac{r}{v_{\mathrm{tr}} t} \right)^{- \delta}, & v_{\mathrm{ej}, \min}
    \leqslant v < v_{\mathrm{tr}}\\
    \zeta_{\rho} \frac{M_{\mathrm{ej}}}{v_{\mathrm{tr}}^3 t^3} \left(
    \frac{r}{v_{\mathrm{tr}} t} \right)^{- n}, & v_{\mathrm{tr}} \leqslant v
    \leqslant v_{\mathrm{ej}, \max}
  \end{array} \right.
\end{equation}
where the transition velocity $v_{\mathrm{tr}}$ could be obtained from the density continuity condition
\begin{equation}
\begin{aligned}
  v_{\mathrm{tr}} &= \zeta_v \left( \frac{E_{\mathrm{sn}}}{M_{\mathrm{ej}}}
  \right)^{1 / 2} \\ \nonumber
  &\simeq 1.2 \times 10^4 \mathrm{km} \text{ s}^{- 1}  \left(
  \frac{E_{\mathrm{sn}}}{10^{51} \mathrm{erg}} \right)^{1 / 2} \left(
  \frac{M_{\mathrm{ej}}}{M_{\odot}} \right)^{- 1 / 2} .
  \end{aligned}
\end{equation}
The numerical coefficients depend on the density power indices as \citep{kasen16}
\begin{equation}
  \zeta_{\rho} = \frac{(n - 3) (3 - \delta)}{4 \pi (n - \delta)}, \hspace{1em}
  \zeta_v = \left[ \frac{2 (5 - \delta) (n - 5)}{(n - 3) (3 - \delta)}
  \right]^{1 / 2} .
\end{equation}
For core collapse SNe, the typical values of the density power indices are $\delta = 1, n = 10$ \citep{chevalier89}. 

Here we assume that the SN ejecta undergoes a homologous expansion i.e., $r = v t$, where the inner boundary of the ejecta could be defined by the slowest ejecta,
\begin{equation}
  R_{\min} (t) = R_{\min, 0} + v_{\mathrm{ej}, \min} t,
\end{equation}
where $v_{\mathrm{ej}, \min}$ is the minimum velocity of the ejecta and $R_{\min, 0}$ is the initial radius of the innermost radius when the explosion enters the homologous phase. Similarly, the outermost layer of the SN ejecta should be
\begin{equation}
  R_{\max} = R_{\max, 0} + v_{\mathrm{ej}, \max} t,
\end{equation}
where $v_{\mathrm{ej}, \max}$ is the maximum velocity of the ejecta and $R_{\max, 0}$ is the initial radius of the outermost radius in the homologous phase. 

With the expansion, a good fraction of the SN material would enter and be trapped by the gravitational potential of the companion BH. If one takes a spherical coordinate with its origin at the center of the BH, the gravitational binding energy for an SN ejecta element with mass $m$ would be  $E_{\mathrm{gra}} = G M_{\mathrm{BH}} m / r_{}$. If this gravitational binding energy is larger than the kinetic energy of this element $\frac{1}{2}m v^2$, it will be trapped and accreted by the central BH. We thus define an accretion radius of the BH as 
\begin{equation}
  R_{\mathrm{acc}} = \frac{2 G M_{\mathrm{BH}}}{v^2} \simeq 5.3 \times 10^9
  \mathrm{cm} \left( \frac{M_{ \mathrm{BH}}}{20 M_{\odot}} \right) \left(
  \frac{v}{10^4 \mathrm{km}~\text{s}^{- 1}} \right)^{- 2}.
\end{equation}
When the outermost radius of SN ejecta reaches $d - R_{\mathrm{acc}}$, the outer part of the SN ejecta with $\rho_{\mathrm{ej}} \propto r^{- n}$ begins to fall into the BH.\quad The time for the falling \ process is set as
\begin{equation}
\begin{aligned}
  t_{\mathrm{start}} &= \frac{d - R_{\mathrm{acc}} - R_{\max, 0}}{v_{\mathrm{ej},
  \max}} \\ 
  &\sim 10^{4} ~{\rm s} \left(\frac{d}{10^{13}{\rm cm}}\right)\left(\frac{v_{\mathrm{ej}, \max}}{10^4 \mathrm{km}~\text{s}^{- 1}}\right)^{-1} .
  \end{aligned}
\end{equation}
In this phase, the material falling rate is
\begin{equation}
\begin{aligned}
  \dot{M} & \simeq \pi \text{} R_{\mathrm{acc}}^2 v \rho_{\mathrm{ej}}, \\ 
  &= \frac{4 \pi G^2 M_{\mathrm{BH}}^2}{d^3} \zeta_{\rho}
  \frac{M_{\mathrm{ej}}}{v_{\mathrm{tr}}^3} \left( \frac{d}{v_{\mathrm{tr}} t}
  \right)^{- n}, \quad t_{\mathrm{start}} \leqslant t < t_{\mathrm{tr}} ,
  \end{aligned}
\end{equation}
where $t_{\mathrm{tr}} \sim d / v_{\mathrm{tr}}$ is the characteristic time when the falling region reaches down to the inner part of the  ejecta, namely when the velocity of falling ejecta element $v$ becomes the transition velocity $v_{\mathrm{tr}}$. The falling rate at the characteristic time is
\begin{equation}
\begin{aligned}
  \dot{M}_{\mathrm{tr}} \simeq &4.1 \times 10^{- 9} M_{\odot}~\text{s}^{- 1}
  \left( \frac{M_{\mathrm{ej}}}{10M_{\odot}} \right)^{5 / 2} \left(
  \frac{M_{\mathrm{BH}}}{20 M_{\odot}} \right)^2 \times\\ 
  &\times\left( \frac{d}{10^{13}
  \mathrm{cm}} \right)^{- 3} \left( \frac{E_{\mathrm{sn}}}{10^{51} \mathrm{erg}}
  \right)^{- 3 / 2}.\label{mff}
  \end{aligned}
\end{equation}
When $t > t_{\mathrm{tr}}$, the density structure of the SN ejecta falling into the BH starts to follow $\rho_{\mathrm{ej}} \propto r^{- \delta}$, so that the falling rate becomes
\begin{equation}
  \dot{M} = \dot{M}_{\mathrm{tr}} \left( \frac{t}{t_{\mathrm{tr}}} \right)^{
  \delta}, \quad t_{\mathrm{tr}} \leqslant t \leqslant t_{\mathrm{end}},
\end{equation}
where $t_{\mathrm{end}} \sim d / v_{\mathrm{ej}, \min}$ is taken as the termination timescale of the falling process. After $t_{\mathrm{end}}$, the accretion rate should not have a 	sudden stop, since materials that are marginally bound to the BH will continue outward on an eccentric orbit and eventually fall back into the BH. Considering that there is no longer a steady influx of material after $t_{\mathrm{end}}$ and the feedback from the disk may blow away some of the marginally bounded materials, here we treat the tail of the accretion rate with an exponential cutoff as $\dot{M}=\dot{M}_{\mathrm{tr}}(t_{\mathrm{end}}/t_{\mathrm{tr}})^{\delta}e^{-(t-t_{\mathrm{end}})/t_{\mathrm{end}}}$.The accretion timescale could be estimated as $t_{\rm acc}\sim t_{\rm ff}/\alpha$, where $\alpha\sim 0.1-0.01$ is the standard dimensionless viscosity parameter \citep{shakura73} and $t_{\rm ff}$ is the timescale for material freely falling from $R_{\mathrm{acc}}$ to the BH, which is
\begin{eqnarray}
t_{\rm ff}&=&\left(\frac{3\pi}{32G}\frac{\frac{4\pi}{3}R_{\mathrm{acc}}^3}{M_{\rm BH}}\right)^{1/2} \nonumber \\ 
&=&0.68~{\rm s}\left(\frac{R_{\mathrm{acc}}}{10^{9}{\rm cm}}\right)^{3/2}\left(\frac{M_{\rm BH}}{20 M_{\odot}}\right)^{-1/2}.
\end{eqnarray}
Obviously, the accretion timescale is much smaller than the dynamical timescale. We thus take the fast accretion approximation and assume the BH accretion rate $\dot{M}_{\rm acc}$ roughly equals to the falling rate $\dot{M}$. Note that some falling materials with small intercept between their motion direction and the BH, may not have sufficient angular momentum to form an accretion disk, but rather fall into the black hole in a roughly spherical fashion. According to the analytical results provided in \cite{kumar08}, materials moving with an intercept of $R'$ relative to the BH could fall to an accretion orbit with $r_{\rm{orb}}\approx R'\left(\Omega/\Omega_{k}\right)^{2}$, where $\Omega$ is angular velocity of the material, $\Omega_{k}$ is the local Keplerian angular velocity. $R'$ with relevant $r_{\rm{orb}}$ equaling to the marginally stable orbit radius $R_{\rm ms}$ (see definition in \ref{rms}) could be defined as the cross section radius for no-disk-formation falling, which could be estimated as $1/R'^{2}=v^{2}/GM_{\rm{BH}}R_{\rm ms}-1/d^{2}$. For the parameter space used in this work,  $R'$ is much smaller than $R_{\rm acc}$. We thus ignore this effect in the following calculations. For cases with extremely slow ejecta velocity or with extremely small orbital separation, the accretion rate could be largely reduced, which is only a fraction of $(R_{\rm acc}^2-R'^2)/R_{\rm acc}^2$ of the falling rate.

According to Eq. \ref{mff}, the accretion process is super-Eddington. In this case, the accretion process could have strong feedback to the SN explosion. Here we consider three feedback mechanisms: 1) accretion disk radiation; 2) Blandford-Znajek jet \citep{BZ}; and 3) Blandford-Payne outflow \citep{BP}. 

We treat the disk evolution as a multicolor blackbody, then the effective temperature of the disk is \citep{strubbe09}
\begin{equation}
  \sigma T_{\mathrm{eff}}^4 = \frac{3 G M_{\mathrm{BH}} \dot{M} f}{8 \pi R^3}
  \times \left[ \frac{1}{2} + \left\{ \frac{1}{4} + \frac{3}{2} f \left(
  \frac{10 \dot{M}}{\dot{M}_{\mathrm{Edd}}} \right)^2 \left( \frac{R}{R_{\rm S}}
  \right)^{- 2} \right\}^{1 / 2} \right]^{- 1},
\end{equation}
where $f= 1 - (R_{\mathrm{ms}} / R)^{1 / 2}, R_{\rm S} = 2 R_{\rm g}$, and $r_{\rm g} = G M_{\rm BH}/c^2$. $ R_{\mathrm{ms}} $ is the marginally stable orbit radius in units of $r_{\rm g}$, and is expressed as \citep{bardeen72,page74}
\begin{equation}
  R_{\mathrm{ms}} = 3 + Z_2 - [(3 - Z_1) (3 + Z_1 + 2 Z_2)]^{1 / 2},
  \label{rms}
\end{equation}
where $Z_1 \equiv 1 + (1 - a^2)^{1 / 3} [(1 + a)^{1 / 3} + (1 - a)^{1 / 3}],
\quad Z_2 \equiv (3 a^2 + Z_1^2)^{1 / 2}$. Here, $a = J_{\rm BH} c/(GM_{\rm BH}^2)$ is the BH spin parameter. The disk luminosity is thus given by
\begin{equation}
  L_{\mathrm{disk}} = 2 \int_{R_{\mathrm{ms}}}^{R_{\mathrm{out}}} 2 \pi R \sigma
  T_{\mathrm{eff}}^4 d R.
\end{equation}
In our case (super-Eddington accretion), we find that $L_{\mathrm{disk}} \sim 0.2 L_{\mathrm{Edd}}\sim 5\times10^{38}{\rm erg~s^{-1}}(M_{\rm BH}/20M_\odot)$. 

The BZ jet power could be estimated as \citep{lee00,li00,wang02,mckinney05,lei11,lei13,liu17}
\begin{equation}
L_{\rm BZ}=1.7 \times 10^{50} a^2 \left(\frac{M_{\rm BH}}{M_{\odot}}\right)^2
B_{\rm H,15}^2 F(a) \ {\rm erg \ s^{-1}},
\label{eq_Lmag}
\end{equation}
where $F(a)=[(1+q^2)/q^2][(q+1/q) \arctan q-1]$ with $q= a /(1+\sqrt{1-a^2})$. 
$B_{\rm H}$ is the magnetic field strength threading the BH horizon, which could be estimated by equating the magnetic pressure on the horizon to the ram pressure of the accretion flow at its inner edge \cite[e.g.][]{moderski97},
\begin{equation}
\frac{B_{\rm H}^2}{8\pi} = P_{\rm ram} \sim \rho c^2 \sim \frac{\dot{M} _{\rm acc}c}{4\pi r_{\rm H}^2},
\label{Bmdot}
\end{equation}
where $r_{\rm H}=(1+\sqrt{1-a^2})r_{\rm g}$ is the radius of the BH horizon. In our case, the BZ jet luminosity could be written as 
\begin{equation}
  L_{\mathrm{BZ}} (t) = \eta_{\mathrm{BZ}} \dot{M}_{\mathrm{tr}} c^2 \left\{
  \begin{array}{ll}
    \left( \frac{t}{t_{\mathrm{tr}}} \right)^{10}, & t_{\mathrm{start}} \leqslant
    t < t_{\mathrm{tr}}\\
    \left( \frac{t}{t_{\mathrm{tr}}} \right), & t_{\mathrm{tr}} \leqslant t
    \leqslant t_{\mathrm{end}}\\
    \left( \frac{t_{\rm end}}{t_{\mathrm{tr}}} \right)e^{-\frac{t-t_{\mathbf{end}}}{t_{\mathbf{end}}}}, &  t
    > t_{\mathrm{end}}
  \end{array} \right.
\end{equation}
where $\eta_{\mathrm{BZ}}=0.52 a^2 F(a)/(1+\sqrt{1-a^2})^2$, we have $\eta_{\mathrm{BZ}}=0.0008$ for $a=0.1$, and $\eta_{\mathrm{BZ}}=0.17$ for $a=0.9$. When the SN expands to a radius of $R_{\rm SN}$, it will roughly take $t_{B}\sim 3000~s\times L_{\rm BZ,45}^{-1/3}\theta_{10^{\rm o}}^{4/3}R_{13}^{2/3}M_{10\odot}^{1/3}$ for the BZ jet to breakout the SN material \citep{bromberg11}. Since the breakout timescale is smaller than the termination timescale of the accretion process, the BZ jet very likely penetrates through the SN envelope. In this case, most of the BZ jet power would dissipate outside of the SN instead of injecting energy into the SN material. Therefore, the feedback effect from BZ power could be neglected here. 

On the other hand, the BP outflow luminosity could be estimated as \citep{armitage99}
\begin{equation}
  L_{\mathrm{BP}} = \frac{(B^P_{\mathrm{ms}})^2 r_{\mathrm{ms}}^4
  \Omega_{\mathrm{ms}}^2}{32 c},
\end{equation}
where $r_{\mathrm{ms}} = R_{\mathrm{ms}} r_{\rm g}$ is the marginally stable orbit radius. $\Omega_{\mathrm{ms}}$ is the Keplerian angular velocity at the marginally stable orbit radius, which could be calculated as 
\begin{equation}
  \Omega_{\mathrm{ms}} = \frac{c^3}{G M_{\mathrm{BH}}} \frac{1}{(R_{\mathrm{ms}}^{3
  / 2} + a)}.
\end{equation}
The poloidal disk magnetic field $B_{\mathrm{ms}}^P$ \ has a relationship with the magnetic field strength threading the BH horizon $B_{\rm H}$ as \citep{BP}
\begin{equation}
  B_{\mathrm{ms}}^P = B_{\rm H} \left( \frac{r_{\mathrm{ms}}}{r_{\rm H}} \right)^{- 5 / 4}.
\end{equation}
In our case, we can derive the BP outflow luminosity as
\begin{equation}
  L_{\mathrm{BP}} (t) = \eta_{\mathrm{BP}} \dot{M}_{\mathrm{tr}} c^2 \left\{
  \begin{array}{ll}
    \left( \frac{t}{t_{\mathrm{tr}}} \right)^{10}, & t_{\mathrm{start}} \leqslant
    t < t_{\mathrm{tr}}\\
    \left( \frac{t}{t_{\mathrm{tr}}} \right), & t_{\mathrm{tr}} \leqslant t
    \leqslant t_{\mathrm{end}}\\
    \left( \frac{t_{\rm end}}{t_{\mathrm{tr}}} \right)e^{-\frac{t-t_{\mathbf{end}}}{t_{\mathbf{end}}}}, &  t
    > t_{\mathrm{end}}
  \end{array} \right.
\end{equation}
where $\eta_{\mathrm{BP}}$ is efficiency, which depends on the BH spin parameter
\begin{equation}
  \eta_{\mathrm{BP}} \equiv \frac{1}{16} \left( 1 + \sqrt{1 - a^2} \right)^{1 /
  2} \frac{R_{\mathrm{ms}}^{3 / 2}}{(R_{\mathrm{ms}}^{3 / 2} + a)^2}.
\end{equation}
We have $\eta_{\mathrm{BP}} = 0.006$ for $a = 0.1$ and $\eta_{\mathrm{BP}} = 0.013$ for $a = 0.9$. Comparing with the BZ jet, BP outflow is less collimated, therefore most of the BP power could be injected into the SN envelope. 

In this scenario, the SN bolometric luminosity can be expressed by \cite{arnett82}
\begin{equation}
  L_{\mathrm{SN}} (t) = e^{- \left( \frac{t^2}{\tau_m^2} \right)} \int_0^t 2
  \frac{t}{\tau_m^2} L_{\mathrm{heat}} (t') e^{\left( \frac{t^{\prime
  2}}{\tau_m^2} \right)} d t'
\end{equation}
where $\tau_m$ is the effective diffusion timescale,
\begin{equation}
  \tau_m = \left( \frac{2 \kappa M_{\mathrm{ej}}}{\beta v c} \right)^{1 / 2}
\end{equation}
where $\kappa$ is the opacity of the SN ejecta, $\beta = 13.8$ is a constant for the density distribution of the ejecta. Here we take \footnote{For cases we are interested in, $L_{\mathrm{BP}}$ is always larger than $L_{\mathrm{disk}}$.}
\begin{equation}
  L_{\mathrm{heat}} (t) = L_{\mathrm{disk}} (t) + L_{\mathrm{BP}} (t) + L_{\mathrm{Ni}} (t),
\end{equation}
where $L_{\mathrm{Ni}}$ is the heating power from the radioactive decay of $^{56}$Ni. 

\section{Results}

Depending on the orbital separation of the binary system $d$ and the SN properties, such as the ejecta mass $M_{\rm ej}$ and the  explosion energy $E_{\mathrm{sn}}$, the accretion feedback power $L_{\mathrm{disk}} + L_{\mathrm{BP}} $ could be larger, comparable or smaller than the radioactive decay heating power. For the last case, the SN lightcurve would behave as a normal core collapse SN. It is very difficult to justify the existence of a companion BH. But for the first two cases, the SN lightcurve could be significantly altered. 

For instance, when the accretion feedback power is much larger than the radioactive heating power, we find that the SN lightcurve would show a sharp peak, whose luminosity could reach the order of $10^{44} \rm erg~s^{-1}$, as luminous as the super luminous SNe \cite[][for a review]{galyam2019}. Here we show an example in Figure 1, where $d=10^{13}$ cm, $M_{\rm ej}=5 M_{\odot}$, $E_{\rm {sn}}=10^{51}$ erg, $v_{\min} =50 \text{ km s}^{-1}$, $M_{\rm BH}=20 M_{\odot}$, $a=0.5$, and $M_{\rm Ni}=0.5 M_{\odot}$ are adopted. In the literature, a newly formed magnetar is commonly proposed to be the energy source of SLSNe \citep{kasen10,woosely10}.  For comparison, we also plot the SN lightcurve when the heating power is dominated by a magnetar with spin period $P=4.5$ ms, and dipole magnetic field $B=2 \times 10^{14}$ G. The lightcurve of our model and the magnetar model are clearly different. For our model, the accretion feedback would terminate when the inner boundary of the ejecta passes over the BH, so that the SN lightcurve would undergo a rapid decay after the peak, and then change to the normal decay as powered by the radioactive decay. But for the magnetar model, the energy injection always continues, so that the SN lightcurve is always dominated by magnetar power, which would undergo a relatively slow decay after the peak. Such different lightcurve behaviors could help us distinguish whether the SLSN is powered by our model or the magnetar model.

On the other hand, when the accretion feedback power is comparable to the radioactive heating power, we find that the SN lightcurve would show a plateau feature. Here we also show an example for this case in Figure 1, where $d=3\times 10^{13}$ cm, $M_{\rm ej}=5 M_{\odot}$, $E_{\rm {sn}}=10^{51}$ erg, $v_{\min} =50 \text{ km s}^{-1}$, $M_{\rm BH}=20 M_{\odot}$, $a=0.5$, and $M_{\rm Ni}=0.5 M_{\odot}$ are adopted. For comparison, we also plot the SN lightcurve when the heating power is dominated by a magnetar with spin period $P=7.5$ ms, and dipole magnetic field $B=10^{14}$ G. In this case, the SN lightcurve of our model would also undergo a rapid decay after the plateau feature, which is clearly distinct from the magnetar model. 

\begin{figure}[tbph]
\begin{center}
\includegraphics[width=0.49\textwidth,angle=0]{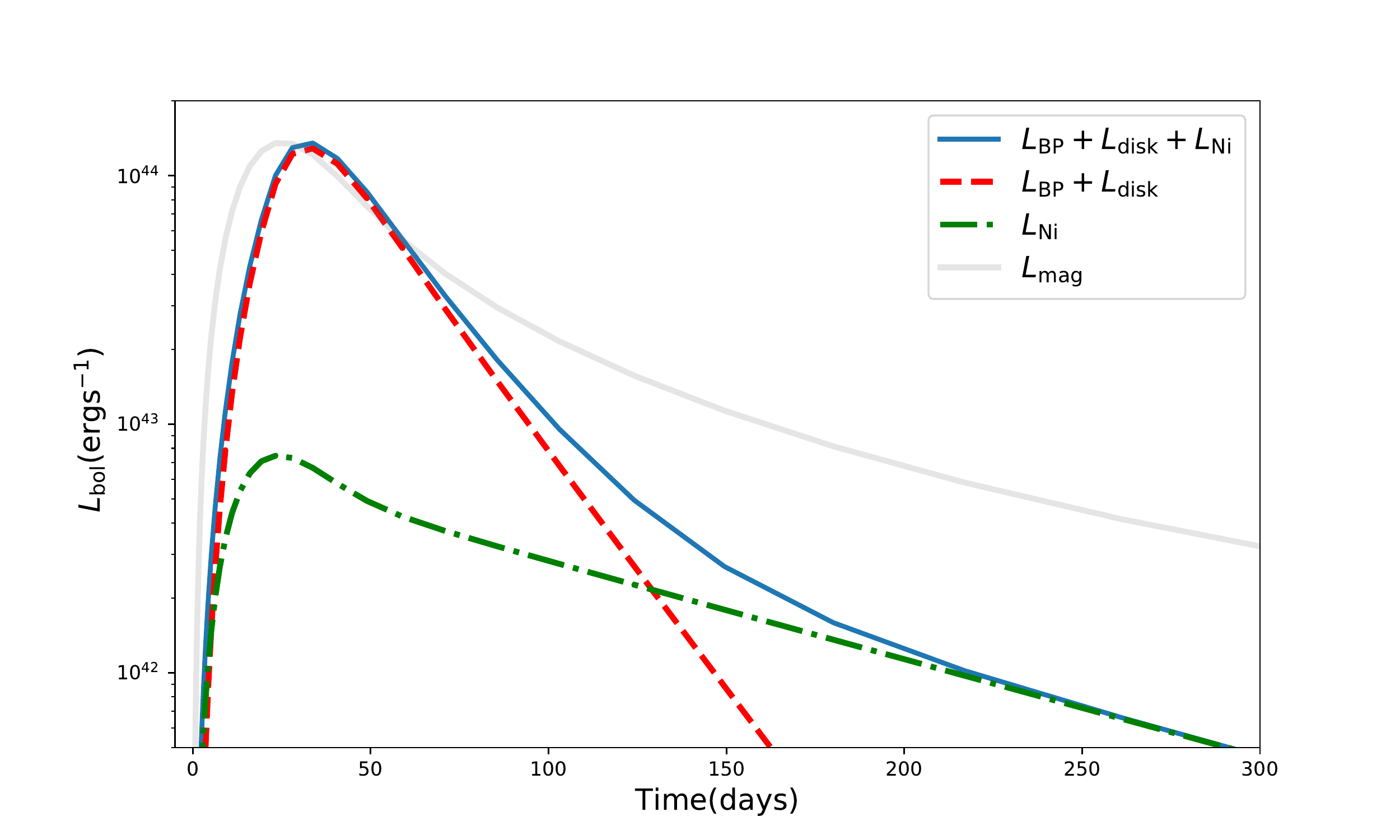}
\includegraphics[width=0.49\textwidth,angle=0]{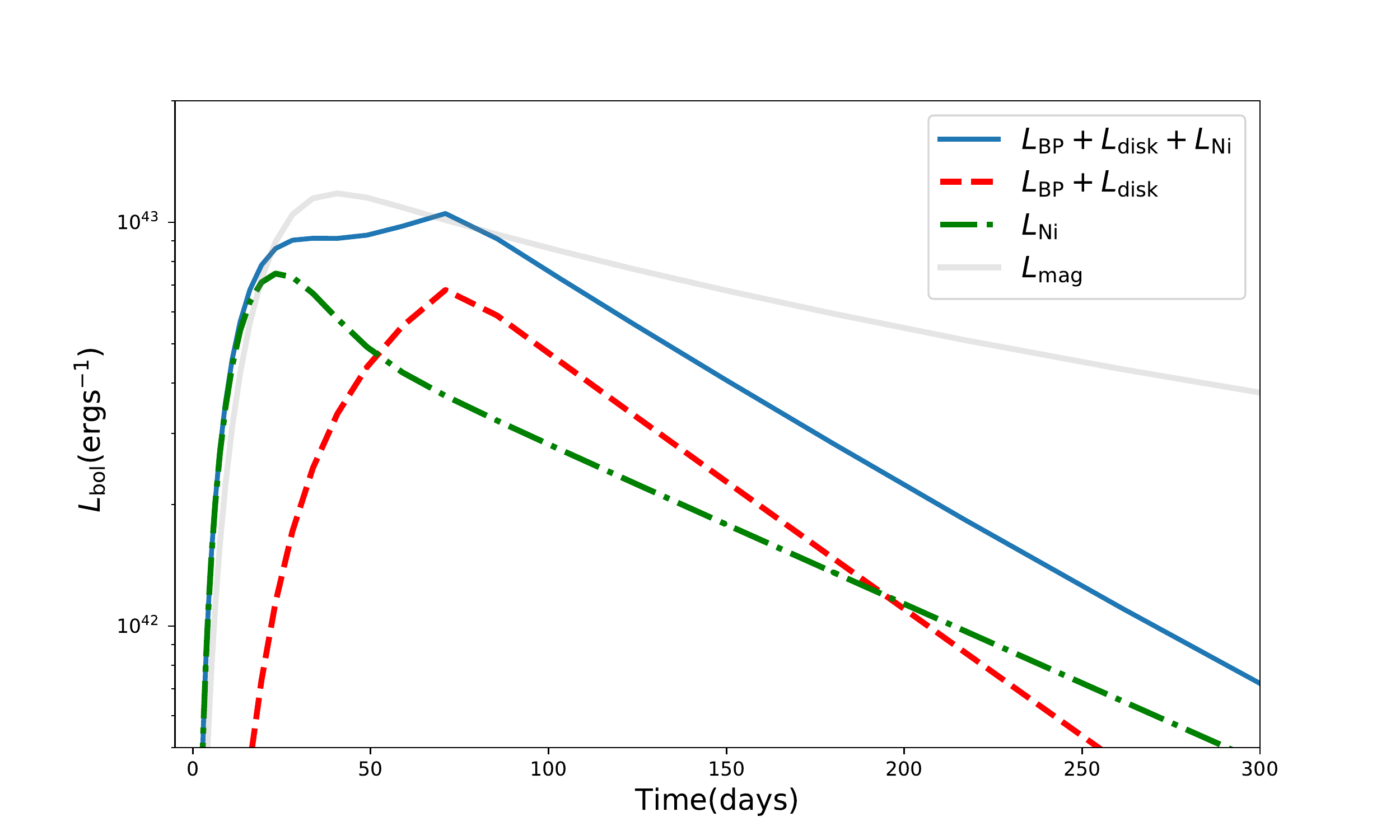}
\end{center}
\caption{Examples of SNe lightcurves the accretion feedback power from the companion BH is larger (upper penal) or comparable to (lower panel) the radioactive heating power.}
\label{fig:lightcurve}
\end{figure}

\section{Conclusion and Discussion}

In this Letter, we propose that if the BH-NS/BH systems detected by LVC are formed from isolated binary evolution, the SN signal associated with the second core collapse would show some identifiable features, due to the accretion feedback from the companion BH. When the feedback power is much greater than the radioactive decay power, the SN lightcurve could show a sharp peak as luminous as the SLSNe (e.g. $\sim10^{44}~\rm erg~s^{-1}$). When the feedback power is comparable to the radioactive decay power, the SN lightcurve could contain a plateau feature. Finally, if the feedback power is much smaller than the radioactive decay, no new features could show up. Note that when the second core collapse leads to a massive BH, it is expected that the SNe ejecta mass would be relatively low, so that the special SNe feature discussed in this work would be more significant in BH-NS progenitor systems or BH-BH progenitor systems where the secondly produced BH having a relatively low mass (e.g. BH-BH systems with asymmetric masses). In some cases, the second core collapse may not even make a successful SNe explosion, but only eject a small fraction of outer envelop. In these cases, a super-Eddington accretion process from the companion BH could still happen, and a powerful disk wind can lead to optical transients with a duration of a few days, and an absolute magnitude ranging from about $-11$ to $-14$ \citep{kimura17}. 

In addition, the accretion of the companion BH is likely to produce a jet via the BZ mechanism \footnote{Note that some hydrodynamical simulations have been performed to explore the jet formation process when an NS is the companion to a type Ic or type Ib core collapse SN \citep{Akashi20}}. The jet would pass through the SN envelope and produce X-ray radiation through internal dissipation and multiband afterglow radiation through external dissipation. These signals will be the direct evidence to identify whether an SN signal is related to the BH-NS/BH system. However, due to the beaming effect of the jet, the radiation can only be seen within the jet opening angle. In the future, with the development of the X-ray and optical sky survey project, the probability of jointly detecting a jet related signal together with the SN signal will be greatly increased. 

In principle, each binary evolution formed BH-NS/BH system would be associated with such an SN signal that we propose here, but only a small fraction could be identified, depending on the modification degree of the lightcurve, and essentially depending on the properties of the binary system, such as the orbital separation $d$. For the examples shown in Figure 1, the orbital separation $d$ needs to be smaller than $3.5 \times 10^{13}$ cm, in order to make the companion BH feedback larger than the radioactive decay power. From the observational  perspective, LVC gives a rough estimation for the event rate density of BBH mergers, e.g., $R=53.2_{-28.2}^{+55.8}~{\rm Gpc^{-3}~yr^{-1}}$\citep{GWT1}. LVC detected BBH systems normally have two black holes with several tens of solar masses. According to the simplest estimation, the initial orbital separation for these systems should be smaller than $1.14 \times 10^{13} {\rm cm}$, otherwise the merger delay time would be larger than the Hubble time. Based on this, we can roughly estimate that the event rate density for our proposed special SN signals would be larger than $53.2_{-28.2}^{+55.8}~{\rm Gpc^{-3}~yr^{-1}}$. In future works, systematically searching for these signals from the SNe archive data to provide their event rate, would be helpful to justify whether the LVC detected BH-NS/BH systems is indeed originated from binary evolution channel.

In the end, we would like to point out some uncertainties for our current results, which would require future complex numerical simulations to fully address. For instance, when the innermost ejecta go beyond the companion BH, material that is marginally bound to the BH will continue outward on an eccentric orbit and eventually fall back into the BH. Here we adopt an exponential cutoff to describe the tail of the accretion rate. If, however, the end of the accretion is slower, the decay phase of the relevant sharp peak or plateau feature would become shallower. On the other hand, as the accretion rate increases, strong feedback from the disk may blow away the loosely bound material, which may halt the accretion process for a certain period. This intermittent accretion process could alter the SNe lightcurve with some oscillation features. Finally, if there is additional energy injection from the second SN produced central remnant (magnetar wind injection if the second remnant is a NS or fallback accretion injection if the second remnant is a BH), the SN lightcurve behavior would become even more complex, where multiple segments or even multiple peaks may show up\footnote{If the explosion of a signal massive star is a partial failure, fallback accretion onto its central remnant could also alter the SN light curve. The main difference is that for single star case, the weaker the initial SN explosion, the stronger the fallback feedback, which is opposite for the scenario discussed here. Ejecta mass and initial velocity measurement could help to distinguish these two progenitors.}.

\acknowledgments
We thank Bing Zhang for helpful discussion and the anonymous referee for the helpful comments that have helped us to improve the presentation of the paper. This work is supported by the grant Nos. 11722324, 11690024, 11703001, 11773010, U2038107, and U1931203 from the National Natural Science Foundation of China, the Strategic Priority Research Program of the Chinese Academy of Sciences, grant No. XDB23040100  and the Fundamental Research Funds for the Central Universities. L.D.L is supported by the National Postdoctoral Program for Innovative Talents (grant No. BX20190044), China Postdoctoral Science Foundation (grant No. 2019M660515)  and ``LiYun'' postdoctoral fellow of Beijing Normal University.

\bibliographystyle{apj} 

\begin{thebibliography}{}
\expandafter\ifx\csname natexlab\endcsname\relax\def\natexlab#1{#1}\fi
\bibitem[Abbott et al.(2016)]{GW150914} Abbott, B.~P., Abbott, R., Abbott, T.~D., et al.\ 2016, \prl, 116, 061102
\bibitem[Abbott et al.(2019)]{GWT1} Abbott, B.~P., Abbott, R., Abbott, T.~D., et al.\ 2019, \apjl, 882, L24
\bibitem[Abbott et al.(2020a)]{GW190412} Abbott, R., et al.\ 2020, arXiv:2004.08342
\bibitem[Abbott et al.(2020b)]{GW190814} Abbott, R., Abbott, T.~D., Abraham, S., et al.\ 2020, \apjl, 896, L44
\bibitem[Anand et al.(2020)]{Anand20} Anand, S., Coughlin, M.~W., Kasliwal, M.~M., et al.\ 2020, Nature Astronomy, doi:10.1038/s41550-020-1183-3
\bibitem[Akashi \& Soker(2020)]{Akashi20} Akashi, M. \& Soker, N.\ 2020, arXiv:2007.07819
\bibitem[Armitage \& Natarajan(1999)]{armitage99} Armitage, P.~J. \& Natarajan, P.\ 1999, \apjl, 523, L7
\bibitem[Arnett(1982)]{arnett82} Arnett, W.~D.\ 1982, \apj, 253, 785
\bibitem[Bardeen et al.(1972)]{bardeen72} Bardeen, J.~M., Press, W.~H., \& Teukolsky, S.~A.\ 1972, \apj, 178, 347
\bibitem[Belczynski et al.(2016)]{belczynski16} Belczynski, K., Repetto, S., Holz, D.~E., et al.\ 2016, \apj, 819, 108
\bibitem[{Blandford \& Znajek}(1977)]{BZ}Blandford, R. D., \& Znajek, R. L., 1977, MNRAS, 179, 433
\bibitem[{Blandford \& Payne}(1982)]{BP}Blandford, R. D., \& Payne, D. G., 1982, MNRAS, 199, 883
\bibitem[Bromberg et al.(2011)]{bromberg11} Bromberg, O., Nakar, E., \& Piran, T.\ 2011, \apjl, 739, L55
\bibitem[Chevalier \& Soker(1989)]{chevalier89} Chevalier, R.~A. \& Soker, N.\ 1989, \apj, 341, 867
\bibitem[Gal-Yam(2019)]{galyam2019} Gal-Yam, A.\ 2019, \araa, 57, 305
\bibitem[Kasen \& Bildsten(2010)]{kasen10} Kasen, D. \& Bildsten, L.\ 2010, \apj, 717, 245
\bibitem[Kasen et al.(2016)]{kasen16} Kasen, D., Metzger, B.~D., \& Bildsten, L.\ 2016, \apj, 821, 36
\bibitem[Kimura et al.(2017)]{kimura17} Kimura, S.~S., Murase, K., \& M{\'e}sz{\'a}ros, P.\ 2017, \apj, 851, 53
\bibitem[Kumar et al.(2008)]{kumar08} Kumar, P., Narayan, R., \& Johnson, J.~L.\ 2008, \mnras, 388, 1729
\bibitem[Lipunov et al.(1997)]{lipunov97} Lipunov, V.~M., Postnov, K.~A., \& Prokhorov, M.~E.\ 1997, \mnras, 288, 245
\bibitem[Lee et al.(2000)]{lee00} Lee, H.~K., Wijers, R.~A.~M.~J., \& Brown, G.~E.\ 2000, \physrep, 325, 83
\bibitem[Lei \& Zhang(2011)]{lei11} Lei, W.-H., \& Zhang, B.\ 2011, \apjl, 740, L27
\bibitem[Lei et al.(2013)]{lei13} Lei, W.-H., Zhang, B., \& Liang, E.-W.\ 2013, \apj, 765, 125
\bibitem[Li(2000)]{li00} Li, L.-X.\ 2000, \prd, 61, 084016
\bibitem[Liu et al.(2017)]{liu17} Liu, T., Gu, W.-M., \& Zhang, B.\ 2017, \nar, 79, 1
\bibitem[Matzner \& McKee(1999)]{matzner99} Matzner, C.~D. \& McKee, C.~F.\ 1999, \apj, 510, 379
\bibitem[McKinney(2005)]{mckinney05} McKinney, J.~C.\ 2005, \apjl, 630, L5
\bibitem[Moderski et al.(1997)]{moderski97} Moderski, R., Sikora, M., \& Lasota, J.~P.\ 1997, Relativistic Jets in AGNs, 110
\bibitem[Page \& Thorne(1974)]{page74} Page, D.~N. \& Thorne, K.~S.\ 1974, \apj, 191, 499
\bibitem[Portegies Zwart \& McMillan(2000)]{portegies00} Portegies Zwart, S.~F. \& McMillan, S.~L.~W.\ 2000, \apjl, 528, L17
\bibitem[Rodriguez et al.(2015)]{rodriguez15} Rodriguez, C.~L., Morscher, M., Pattabiraman, B., et al.\ 2015, \prl, 115, 051101
\bibitem[Sigurdsson \& Hernquist(1993)]{sigurdsson93} Sigurdsson, S. \& Hernquist, L.\ 1993, \nat, 364, 423
\bibitem[Shakura \& Sunyaev(1973)]{shakura73} Shakura, N.~I. \& Sunyaev, R.~A.\ 1973, \aap, 500, 33
\bibitem[Strubbe \& Quataert(2009)]{strubbe09} Strubbe, L.~E. \& Quataert, E.\ 2009, \mnras, 400, 2070
\bibitem[Tutukov \& Yungelson(1973)]{tutukov73} Tutukov, A. \& Yungelson, L.\ 1973, Nauchnye Informatsii, 27, 70
\bibitem[Woosley(2010)]{woosely10} Woosley, S.~E.\ 2010, \apjl, 719, L204
\bibitem[Wang et al.(2002)]{wang02} Wang, D.~X., Xiao, K., \& Lei, W.~H.\ 2002, \mnras, 335, 655




\end{thebibliography}

\end{document}